# Materials development by interpretable machine learning

Yuma Iwasaki[1,2]*, Ryoto Sawada[1], Valentin Stanev[3,4], Masahiko Ishida[1], Akihiro Kirihara[1], Yasutomo Omori[1], Hiroko Someya[1], Ichiro Takeuchi[3], Eiji Saitoh[5,6,7], Yorozu Shinichi[1]

[1]Central Research Laboratories, NEC Corporation, Tsukuba 305-8501, Japan

[2]PREST, JST, Saitama 322-0012, Japan

[3]Department of Materials Science and Engineering, University of Maryland, College park, MD 20742, USA

[4]*Center for Nanophysics and Advanced Materials, University of Maryland, College Park, MD 20742, USA*

[5]Institute for Materials Research, Tohoku University, Sendai 908-8577, Japan

[6]WPI Advanced Institute for Materials Research, Tohoku University, Sendai 908-8577, Japan

[7]Advanced Science Research Center, Japan Atomic Energy Agency, Tokai, 319-1195, Japan

Email: y-iwasaki@ih.jp.nec.com

Machine learning technologies are expected to be great tools for scientific discoveries. In particular, materials development (which has brought a lot of innovation by finding new and better functional materials) is one of the most attractive scientific fields. To apply machine learning to actual materials development, collaboration between scientists and machine learning is becoming inevitable. However, such collaboration has been restricted so far due to black box machine learning, in which it is difficult for scientists to interpret the data-driven model from the viewpoint of material science and physics. Here, we show a material development success story that was achieved by good collaboration between scientists and one type of interpretable (explainable) machine learning called factorized asymptotic Bayesian inference hierarchical mixture of experts (FAB/HMEs). Based on material science and physics, we interpreted the data-driven model constructed by the FAB/HMEs, so that we discovered surprising correlation and knowledge about thermoelectric material. Guided by this, we carried out actual material synthesis that led to identification of a novel spin-driven thermoelectric material with the largest thermopower to date.

Recent progresses of material science technologies enable us to obtain a lot of material data in a short time [1-4]. Accordingly, the development of data-processing technologies for such material big data is becoming indispensable. Machine learning technologies are believed to be a great solution due to not only their rapid data analysis [5-8] but also their potential to discover non-trivial knowledge that is not rooted in conventional material theory.

To apply machine learning to actual materials development, collaboration between scientists and machine learning is becoming inevitable. Material scientists often try to understand the inside of the data-driven model to obtain some actionable information for materials development. However, such collaboration has been restricted so far because of the low interpretability in machine learning. For example, it is not easy for a person to understand the inside of data-driven models constructed by a deep neural network [9], where the data-driven model is expressed as the connections between large numbers of perceptrons (neurons). Therefore, the notion of interpretable machine learning (explainable or transparent machine learning), which has not only high prediction accuracy but also high interpretability, has recently seen a resurgence [10-12], especially for scientific discoveries led by machine learning.

Here, we show an actual material development success story by using one type of interpretable machine learning called factorized asymptotic Bayesian inference hierarchical mixture of experts (FAB/HMEs) [13-16] and demonstrate good collaboration between the FAB/HMEs and material scientists. In the material development, the machine learning algorithm must meet the following three requirements, "sparse modelling", "prediction accuracy", and "interpretability", as shown in Figure 1. The material data is often sparse; therefore, the sparse modelling approach, which automatically selects some of only the important descriptors (attributes), is useful.

One of the most popular sparse modelling methods is LASSO, which constructs a linear model with $L_1$ regularization [17]. However, such a linear regression model does not always have high prediction accuracy because material data often includes non-linear relationships such as phase transitions. For high prediction accuracy, non-linear modelling using, for example, a support vector machine (SVM), neural network (NN), or random forest (RF) can construct a flexible model [17]. However, such non-linear modellings do not have enough interpretability. Although these non-linear machine learnings can tell us which descriptors are important for the data-driven model, they do not clarify how the descriptors actually contribute to it. Extracting actionable information from such non-linear models is not so easy.

Interpretable machine learning FAB/HMEs constructs a piecewise sparse linear modelling [13] that meets the three requirements of "sparse modelling", "prediction accuracy", and "interpretability". Therefore, the actionable information from the data-driven model provided by the FAB/HMEs smoothly leads us to discoveries of novel materials.

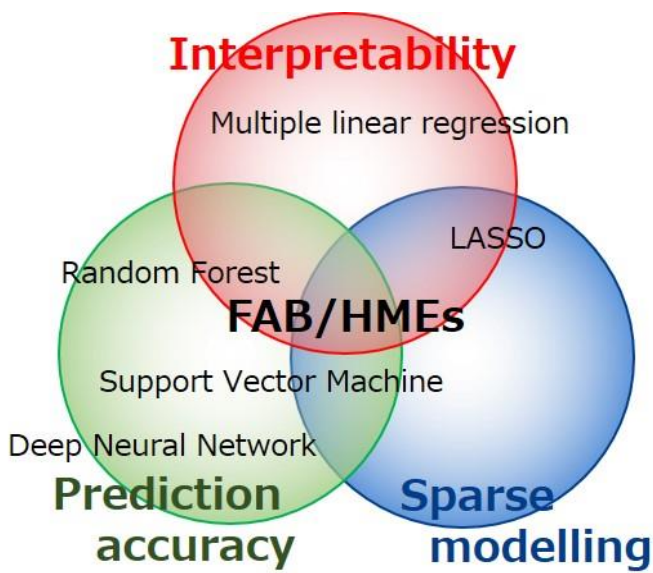

**Figure 1 | Three requirements of machine learning in materials developments.** For good collaboration between machine learning and scientists in materials developments, requirements of sparse modelling, prediction accuracy, and interpretability are

ideal. One type of interpretable machine learning called factorized asymptotic Bayesian inference hierarchical mixture of experts (FAB/HMEs) meets all three requirements.

For the case study, we applied the interpretable machine learning FAB/HMEs to develop a new thermoelectric material. Thermoelectric technologies are becoming indispensable in the quest for a sustainable future [18,19]. In particular, the emerging spin-driven thermoelectric (STE) material has garnered much attention as a promising path toward low cost and versatile thermoelectric technology with easily scalable manufacturing [20-23]. However, STE material development is hampered by the lack of understanding of the fundamental mechanism of STE material because STE phenomena are at the cutting edge of material science and physics. A data-driven approach by machine learning can exhibit its full potential in such an unexplored scientific field.

By searching for STE materials with improved thermopower $S_{STE}$ (one of the most important figures of merit for thermoelectric phenomena) [24], we have used interpretable machine learning FAB/HMEs to discover non-trivial behaviors of material parameters governing the STE phenomena. We have successfully leveraged this machine-learning-informed knowledge to discover a novel high-performance STE material, whose thermopower is far greater than that of the best known STE material to date [22].

**Result**

**Material data.** Figures 2(a-c) show the material data of STE devices using a $M_{100-x}Pt_x$ binary alloy, where M = Fe, Co, and Ni. The thermopower $S_{STE}$ is experimental data on different experimental conditions (different substrates) $\mathbf{C} \equiv \{Si, AlN, GGG\}$ (the details about the experimental conditions are in Supplemental Information 1).

The material parameters $\mathbf{X} \equiv \{X_1, X_2, X_3 ... X_{14}\}$, whose simple descriptions are shown in Table 1, were obtained by conventional material simulation technology (high-throughput DFT calculation [3,4]). The details about the data, experiments, DFT calculations, data-preprocessing, and the reason we use these material parameters $\mathbf{X}$ are shown in Supplemental Information 1.

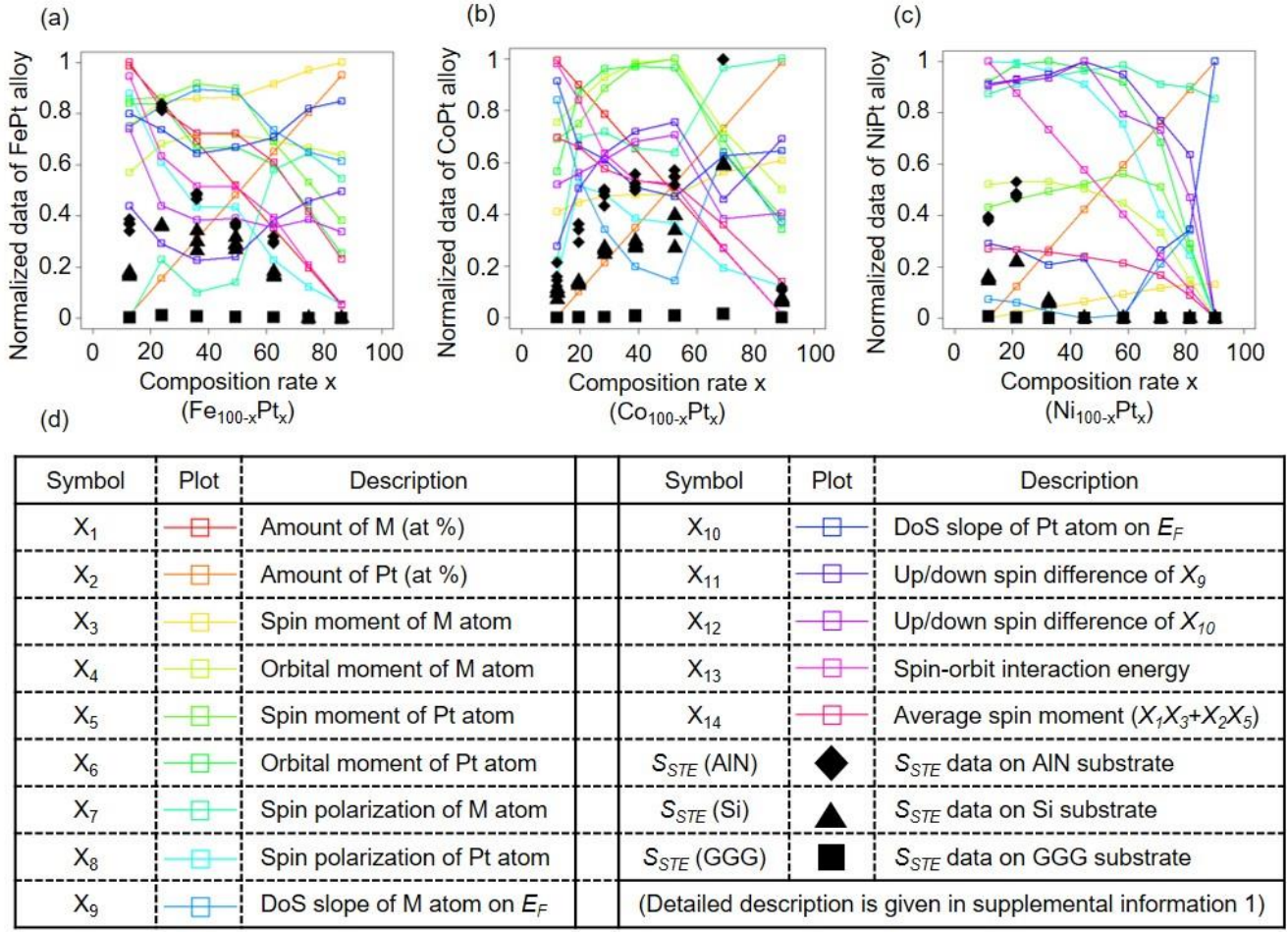

**Figure 2 | STE material data.** Data of spin-driven thermopower $S_{STE}$ and descriptors $X$ used in machine learning modelling. **a.** FePt binary alloy. **b.** CoPt binary alloy. **c.** NiPt binary alloy. **d.** Table of simple description about these STE material data.

**Machine learning modelling by FAB/HMEs.** With these data, we constructed the following data-driven model by the FAB/HMEs.

$$S_{STE} = f(\mathbf{X}, \mathbf{I}, \mathbf{S}, \mathbf{C}) \tag{1}$$

where **X, I, S,** and **C** are the material parameters $X \equiv \{X_1, X_2, X_3 ... X_{14}\}$, their interaction terms $I \equiv \{X_1X_2, X_1X_2, X_1X_3 ... X_{13}X_{14}\}$, their square terms $S \equiv \{X_1^2, X_2^2, X_3^2 ... X_{14}^2\}$, and experimental condition terms $C \equiv \{C_{AlN}, C_{Si}, C_{GGG}\}$, respectively. The experimental condition terms **C** are binary parameters (i.e. $C_{AlN} = \{0,1\}$, $C_{Si} = \{0,1\}$, $C_{GGG} = \{0,1\}$). The details are in Supplemental Information 1). The FAB/HMEs can solve data-classification and data-regression problems at the same time by maximizing a novel information criterion (factorized information criterion, which is referred to as FIC) with an EM-like algorithm (factorized asymptotic Bayesian inference, referred to as FAB), thus constructing a piecewise sparse linear model [13-16].

Figure 3a and 3b show the visualization of the data-driven model constructed by the FAB/HMEs. The data is classified according to the tree structure, as shown in Figure 3a. For each data group, regression models (component 1, 2, 3, and 4, as shown in Figure 3b) are created. Note that the FAB/HMEs does not sequentially carry out the data classification and data regression. The FAB/HMEs searches for proper regression models while searching for proper data groups, thus selecting a better combination of regression models and data groups from a very large model-searching space. The prediction accuracy of this model is comparable with other non-linear machine learning models. The details are given in the Discussion section.

The high interpretability of the FAB/HMEs helps us understand the inside of the data-driven model, where we only have to think about the data groups shown by the tree structure in Figure 3a and the regression models (component 1, 2, 3, and 4) for each data group shown in Figure 3b. Therefore, material scientists can interpret the data-driven model based on the viewpoint of physics and material science.

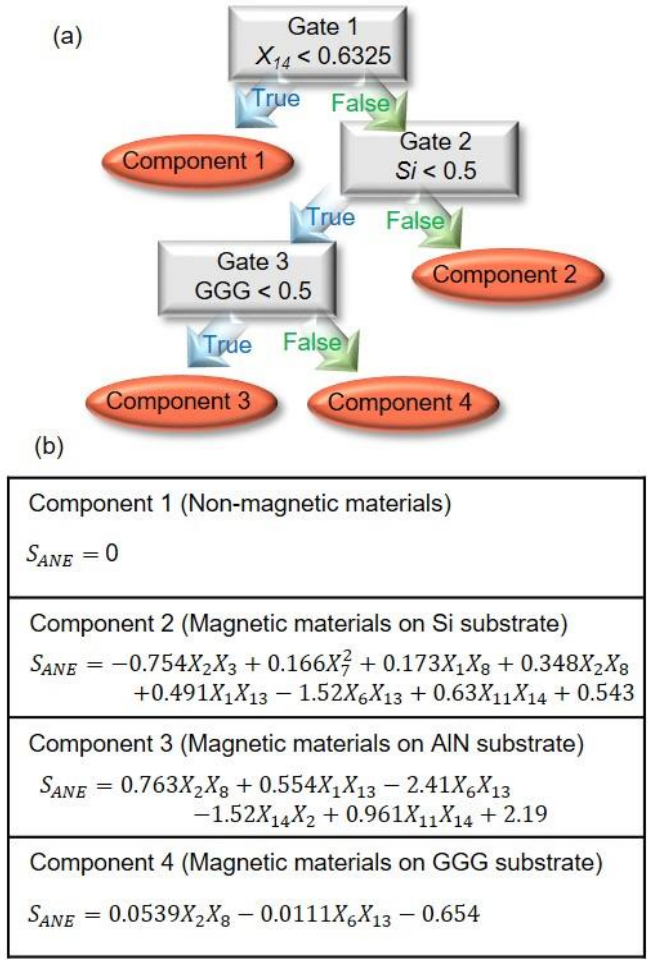

**Figure 3 | Interpretable data-driven model created by FAB/HMEs.** One type of interpretable machine learning called factorized asymptotic Bayesian inference hierarchical mixture of experts (FAB/HMEs) creates piecewise sparse linear model, which is visualized with **a.** tree structure and **b.** regression models. Scientists can interpret tree structure and regression models from viewpoint of material science and physics. Accordingly, notice that there is non-trivial knowledge in data-driven model.

For instance, we notice that the data is firstly classified by average spin moment ($X_{14}$) at gate 1. The data with small average spin moment $X_{14}$ go to component 1, where the thermopower $S_{STE}$ is equal to zero ($S_{STE} = 0$, as shown in Figure 3b). This is natural from the viewpoint of material science and physics. It is known that STE phenomena are not observed on non-magnetic materials with small average spin moment $X_{14}$[25]. The data of magnetic materials with large average spin moment $X_{14}$ only have finite $S_{STE}$ values; accordingly, they go to gate

2. Similarly, from the interpretable data-driven model, we can obtain some trivial knowledge that is consistent with material science and physics (this trivial knowledge is shown in Supplemental Information 2).

Fortunately, we sometimes obtain non-trivial knowledge from the data-driven model. We notice that there is a positive correlation between $S_{STE}$ and product term $X_2X_8$, where $X_2$ and $X_8$ are the amount of Pt atoms and the Pt spin polarization, respectively. All regression models of magnetic materials (component 2, 3, and 4) have the $X_2X_8$ term as positive, as shown in Figure 3b. This positive correlation uncovered by the machine learning models appears to be beyond our current knowledge of STE. The details of the physical interpretation underlying this relation is also discussed in Supplemental Information 2. The surprising connection between $S_{STE}$ and $X_2X_8$, which was discovered by the interpretable machine learning here, will lead to a more comprehensive understanding of the fundamental mechanism of STE phenomena.

**Actual material development guided by FAB/HMEs modelling.** Aside from the profound physical/theoretical discussion about the discovered non-trivial knowledge shown in Supplemental Information 2, we now focus on demonstrating that this unanticipated result of the interpretable machine learning can indeed help us to develop novel STE materials.

It has been difficult to develop STE materials because the fundamental mechanism of STE phenomena, which is at the cutting edge of material science and physics, has not been understood well yet. The materials development in such an unexplored scientific field becomes simplified via the interpretable machine learning modelling. The machine-learning-informed knowledge, which is the positive correlation between $S_{STE}$ and $X_2X_8$, suggests that we just have to search for materials with large $X_2X_8$ to obtain large $S_{STE}$. Searching for a material with large $X_2X_8$ is feasible because we can simulate the $X_2X_8$ value by using conventional material simulation technology.

As a result of material screening for large $X_2X_8$ by the conventional material simulation, we found that $Co_{50}Pt_{50}N_{10}$ has large $X_2X_8$ (the details are shown in Supplemental Information 3). Therefore, as an initial example, we carried out actual material synthesis of $Co_{50}Pt_{50}N_x$ and measured its thermopower $S_{STE}$. Figure 4 shows the results of thermopower $S_{STE}$ of $Co_{50}Pt_{50}N_x$ materials with different N concentration. It is clear that $S_{STE}$ increases with increasing $X_2X_8$ accompanying an increase in N concentration. The $S_{STE}$ of $Co_{45.6}Pt_{47.7}N_{6.7}$ was achieved at 13.04 μV/K, which is larger than that of the current generation of STE materials [22,23]. Details of this actual material synthesis are given in Supplemental Information 4. This data-driven material development succeeded because the interpretable machine learning helped us a great deal by disclosing the non-trivial positive correlation between $S_{STE}$ and $X_2X_8$ due to its high interpretability. This novel STE material with the dramatically enhanced thermopower discovered here will be a key component in thermal energy harvesting systems, leading to a more energy-efficient future.

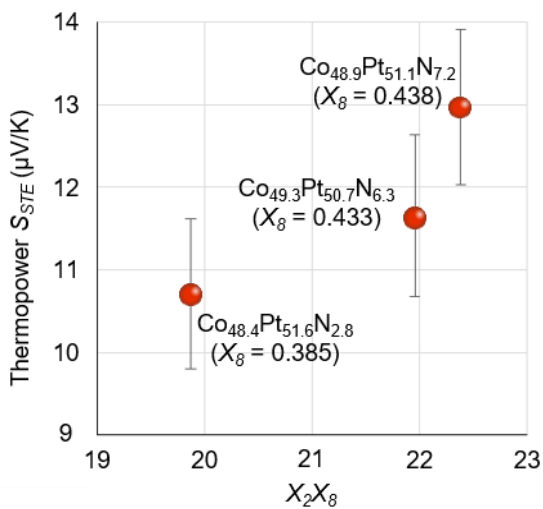

**Figure 4 | Spin-driven thermopower $S_{STE}$ of CoPtN thin films.** $S_{STE}$ increases with increasing $X_2X_8$ in CoPtN, which was selected by material screening guided by FAB/HMEs non-trivial knowledge, positive correlation between $S_{STE}$ and $X_2X_8$. $S_{STE}$ of $Co_{48.9}Pt_{51.1}N_{7.2}$ thin film was achieved at 13.04 μV/K, which is larger than that of STE materials to date.

**Discussion**

Figure 5 shows the 4-fold cross-validation root mean square error (CV-RMSE) of the FAB/HMEs and major machine learning algorithms including the neural network (NN), support vector machine (SVM), random forest (RF), least absolute shrinkage and selection operator (LASSO), and multiple linear regression (MLR). Non-linear models using the FAB/HMEs, NN, SVM, and RF have better prediction accuracy than linear models using LASSO and MLR. The prediction performance of the FAB/HMEs is comparable to that of other non-linear models.

Although the NN and SVM exhibit high prediction performance, it is not easy for scientists to investigate the data-driven model because of their low interpretability. Even though the RF, which tells us the feature importance of each descriptor, has better interpretability than the NN and SVM, further investigation is needed to understand the more specific relationship between $V_{ANE}$ and descriptors. For example, if you only look at the RF model, it is difficult to notice the "positive" correlation between $V_{ANE}$ and $X_2 X_8$ "on the condition of" the magnetic material data (data with large $X_{14}$). Actually, this surprising correlation was not achieved directly via the MLR, LASSO, NN, SVM and RF. On the other hand, the FAB/HMEs obviously visualized such a detailed relationship, as shown in Figure 3a and 3b. As described above, the interpretable machine learning FAB/HMEs provides the data-driven model with not only high prediction performance but also high interpretability even in the case of sparse data.

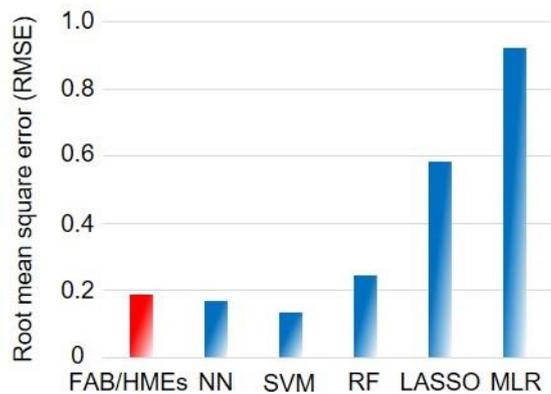

**Figure 5 | Prediction accuracy of FAB/HMEs.** Comparison of 4-fold cross-validation root mean square error (CV-RMSE) of major machine learnings. FAB/HMEs has not only high interpretability but also prediction accuracy that is comparable to other black box machine learning such as NN, SVM, and RF.

In summary, we have demonstrated the utility of interpretable machine learning FAB/HMEs in the material development process. Because of its high prediction performance and interpretability, material scientists can collaborate with the data-driven model obtaining non-trivial knowledge for novel materials development. Guided by the surprising correlation discovered by the FAB/HMEs, we have succeeded in developing an innovative STE material, whose thermopower $S_{STE}$ is much larger than that of the current generation of STE materials. In addition, the non-trivial knowledge we found in the data-driven model can lead to a more comprehensive understanding of the mechanism of emerging STE phenomena. Thus, the enhanced collaboration between scientists and machine learning can lead to not only materials developments but also diverse scientific discoveries.

**Methods**

**FAB/HMEs** [13-16]. The factorized asymptotic Bayesian inference hierarchical mixture of experts (FAB/HMEs) constructs a piecewise sparse linear model that assigns sparse linear experts to individual partitions in feature space and expresses whole models as patches of local experts. By maximizing the factorized information criterion including two $L_0$-regularizations for partition-structure determinations and feature selection for individual experts, the FAB/HMEs performs the partition-structure determination and feature selection at the same time. To maximize the FIC, a factorized asymptotic Bayesian inference (FAB), which combines an expectation-maximization (EM) algorithm with an automatic shrinkage of non-effective experts, is used. In this paper, we set the termination condition $\delta = 10^{-5}$, shrinkage threshold $\varepsilon = 0.062$, and number of initial experts was 32 (i.e. 5-depth symmetric tree). The 4-fold cross-validation root mean square error (CV-RMSE) was 0.188253, as shown in Figure 5.

**NN** [17]. The neural network (NN) models the data by means of a statistical learning algorithm mimicking the brain. Here, we have utilized a simple 3-layer perceptron. The cross-validation with "caret (nnet)" package in the R programming language decides the number r of hidden units $N_H = 8$ and the decay value $D = 3.91 \times 10^{-3}$. The CV-RMSE was 0.169454, as shown in Figure 5.

**SVR** [17]. The support vector regression (SVR) constructs a data-driven model with a kernel method. Here, we have used the radial basis function (RBF) kernel. We set the cost value $C = 16$ and sigma of the RBF $\sigma = 3.125 \times 10^{-2}$, which were decided by the cross-validation with "caret (svmradial)" package in the R programming language. The CV-RMSE was 0.132847, as shown in Figure 5.

**RF** [17]. The random forest (RF) is an ensemble learning method using a multitude of decision trees. The number of trees (*ntree*) and the number of features (*mtry*) were set to 1000 and 120, respectively. We have performed the RF by using the "caret (rf)" package in the R programming language. The CV-RMSE was 0.246274, as shown in Figure 5.

**LASSO** [17]**.** The least absolute shrinkage and selection operator (LASSO) creates a linear regression model with feature selection by using a $L_1$-regularization term. The complexity parameter λ was set to $3.052 \times 10^{-4}$, which was decided by the cross-validation with the "caret (glmnet)" package in the R programming language. The CV-RMSE was 0.583379, as shown in Figure 5.

**Data preparation of thermopower $S_{STE}$.** The spin-driven thermoelectric performance (thermopower $S_{STE}$) data were obtained by experiments including material fabrication processes and material characterization processes. The details are shown in Supplemental Information 1.

**Data preparation of descriptors *X*.** The descriptors *X* data were calculated by the conventional material simulation technique, the Korringa-Kohn-Rostoker and coherent-potential approximation (KKR-CPA) [26]. The details are shown in Supplemental Information 1.

**Reference**

1. Koinuma, H. & Takeuchi, I. Combinatorial solid-state chemistry of inorganic materials. Nat. Mater. **3,** 429 (2004)

2. Takeuchi, I. et al. Identification of novel compositions of ferromagnetic shape-memory alloys using composition spreads. Nat. Mater. **2,** 180-184 (2003)

3. Curtarolo, S., Hart, G. L. W., Nardelli, M. B. Buongiorno, Nardelli, M., Mingo, N., Sanvito, S. & Levy, O. The high-throughput highway to computational materials design. Nat. Mater. 12, 191-201 (2013)

4. Nishijima, M. et al. Accelerated discovery of cathode materials with prolonged cycle life for lithium-ion battery. Nat. Commun. **5,** 4553 (2014)

5. Butler, K. T., Davies, D. W., Cartwright, H., Isayev, O. & Walsh, A. Machine learning for molecular and materials science. Nature **559,** 547-555 (2018)

6. Xia, R. & Kais, S. Quantum machine learning for electronic structure calculations. Nat. Commun. **9,** 4195 (2018)

7. Kusne, A. G. et al. On-the-fly machine-learning for high-throughput experiments: search for rare-earth-free permanent magnets. Sci. Rep. **4,** 6367 (4014)

8. Iwasaki, I., Kusne, A. G. & Takeuchi, I. Comparison of dissimilarity measures for cluster analysis of X-ray diffraction data from combinatorial libraries. npj Comput. Mater. **3,** 4 (2017)

9. Dimiduk, D. M., Holm, E. A. & Niezqoda, S. R. Perspectives on the Impact of Machine Learning, Deep Learning and Artificial Intelligence on Materials, Processes and Structures Engineering. Integr Mater Manuf Innov **7,** 157-172 (2018)

10. Samek, W., Wiegand, T. & Müller, K-R. Interpretable Artificial Intelligence: Understanding, Visualizing and Interpreting Deep Learning Models. Preprint at arXiv, 1708.08296 (2018)

11. Yang, S. C. –H. & Shafto, P. Interpretable Artificial Intelligence via Bayesian Teaching. In NIPS (2017)

12. Guidotti, R. et al. A Survey of Methods for Explaining Black Box Models. Preprint at arXiv, 1802.01933 (2018)

13. Eto, R., Fujimaki, R., Morinaga, S. & Tamano, H. Fully-Automatic Bayesian Piecewise Sparse Linear Models. In AISTATS (2014)

14. Fujimaki, R. & Hayashi, K. Factorized Asymptotic Bayesian Hidden Markov Models. In ICML (2012)

15. Fujimaki, R. & Morinaga, S. Factorized Asymptotic Bayesian Inference for Mixture Modeling. In AISTATS (2012)

16. Hayashi, K. & Fujimaki, R. Factorized Asymptotic Bayesian Inference for Latent Feature Models. In NIPS (2013).

17. Bishop, C. M. Pattern Recognition and Machine Leaning (Springer, 2006)

18. Goldsmid, H. J. *Introduction to Thermoelectricity* (Springer, 2010).

19. Bell, L. E. Cooling, heating, generating power, and recovering waste heat with thermoelectric systems. *Science* **321,** 1457–1461 (2008).

20. Kirihara, A. et al. Spin-current-driven thermoelectric coating. Nature mater. **11,** 686-689 (2012)

21. Sakuraba, Y. Potential of thermoelectric power generation using anomalous Nernst effect in magnetic materials. Scr. Mater. **111,** 29 (2016)

22. Ikhlas, M. et al. Large anomalous Nernst effect at room temperature in a chiral antiferromagnet. Nat. Phys. **13,** 1085-1090 (2017)

23. Iwasaki, Y. et al. Machine-learning guided discovery of a new thermoelectric material. Sci. Rep. **9,** 2751 (2019)

24. Uchida, K. et al. Thermoelectric Generation Based on Spin Seebeck Effects. Proc. IEEE **104,** 1946 (2016)

25. Mizuguchi, M., Ohata, S., Uchida, K., Saitoh, E. & Takanashi, K. Anomalous Nernst Effect in an $L1_0$-Ordered Epitaxial FePt Thin Film. Appl. Phys. Express **5,** 093002 (2012)

26. Akai, H. Electronic Structure Ni-Pd Alloys Calculated by the Self-Consistent KKR-CPA Method. J. Phys. Soc. Jpn. **51,** 468-474 (1982)

**Acknowledgements**

This work was supported by JST-PRESTO "Advanced Materials Informatics through Comprehensive Integration among Theoretical, Experimental, Computational and Data-Centric Sciences" (Grant No. JPMJPR17N4), JST-ERATO "Spin Quantum Rectification Project" (Grant No. JPMJER1402), I.T. is supported in part by C-SPIN, one of six centers of STARnet, a Semiconductor Research Corporation program, sponsored by MARCO and DARPA.

**Author contributions**

Y.I., M.I., A.K., H.S. and O.T. designed the experiment, fabricated the samples, and collected all of the data. Y.I., R.S., and E.S. contribute to the theoretical discussion. Y.I., R.S., V.S. and I.T discussed the results of machine learning modeling. Y.I., V.S., I.T. and M.I wrote the manuscript. S.Y. supervised this study. All the authors discussed the results and commented on the manuscript.

**Additional information**

The authors declare no competing interests.